\documentclass[twocolumn,showpacs,preprintnumbers,amsmath,amssymb,pra]{revtex4}

\usepackage{graphicx}% Include figure files
\usepackage{dcolumn}% Align table columns on decimal point
\usepackage{bm}% bold math
\usepackage{amsmath}
\usepackage{multirow}
\usepackage{mathrsfs}
\usepackage{subfigure}
\usepackage[usenames]{color}
\newcommand{\be}{\begin{eqnarray}}
\newcommand{\ee}{\end{eqnarray}}

\begin{document}
\title{Application of Relativistic Coupled-cluster Theory to Electron Impact Excitations of Mg$^+$ in the Plasma Environment}
\author{L. Sharma$^{1}$, B. K. Sahoo$^{2,3}$, P. Malkar$^{1}$ and R. Srivastava$^{1}$ }
\affiliation{
$^1$Department of Physics, Indian Institute of Technology Roorkee, Roorkee 247667, India\\
$^2$Atomic and Molecular Physics Division, Physical Research Laboratory, Navrangpura, Ahmedabad 380009, India\\
$^{3}$State Key Laboratory of Magnetic Resonance and Atomic and Molecular Physics, Wuhan Institute of Physics and Mathematics,
Chinese Academy of Sciences, Wuhan 430071, China}

\date{\today}

\begin{abstract}
A relativistic coupled-cluster (RCC) theory is implemented to study electron impact excitations of atomic species. As a test case, 
the electron impact excitations of the $3s ~ ^2S_{1/2} - 3p ~ ^2P_{1/2;3/2}$ resonance transitions 
are investigated in the singly charged magnesium (Mg$^+$) ion using this theory. Accuracies of wave functions of Mg$^+$ are justified by 
evaluating its attachment energies of the relevant states and compared with the experimental values. The continuum wave function 
of the projectile electron are obtained by solving Dirac equations assuming distortion potential as static potential of the ground 
state of Mg$^+$. Comparison of the calculated electron impact 
excitation differential and total cross-sections with the available measurements are found to be in very good agreements at various 
incident electron energies. Further, calculations are carried out in the plasma environment in the Debye H\"uckel model framework, 
which could be useful in the astrophysics. Influence of plasma strength on the cross-sections as well as linear
polarization of the photon emission in the $3p ~ ^2P_{3/2} - 3s ~ ^2S_{1/2}$ transition is investigated for different incident 
electron energies.
\end{abstract}


\maketitle
\section{Introduction}

Accurate determination of scattering cross-sections of electrons with atomic systems are of immense interest to a wide range of applications
in astrophysics and plasma physics \cite{anil,johnson,amusia,Post,RDressler,Dipti,Badnell}. The challenges in the calculations of scattering cross-sections lie in determining accurate wave
functions for the ground and excited states of the target as well as calculating wave functions of the scattered electron in the vicinity of the target. Furthermore, it is important to evaluate these 
wave functions in the relativistic framework to investigate many subtle effects of the problem. Many-body methods in the close-coupling \cite{CC} 
and R-matrix \cite{Burke} formalisms in the non-relativistic approach, and multi-configuration Dirac-Fock (MCDF) method,  which is a special case 
of a truncated configuration interaction (CI) method, in the relativistic approach are the commonly used \cite{grasp2k} techniques to evaluate the wave functions
of the target atomic systems. However, a truncated CI method does not include the electron correlation effects in correct scaling with the size of
the system \cite{szabo,bartlett}. In contrast, a relativistic coupled-cluster (RCC) method is known today as the golden many-body method in the 
nuclear, atomic and molecular physics for its ability to incorporate the electron correlation effects to all orders in perturbation 
for evaluating wave functions \cite{szabo,bartlett}. A truncated RCC method can still include contributions from higher level excitations even at
the same excitation level approximations in the CI method. Moreover, truncated RCC methods obey the size-extensivity and size-consistent behaviors.
It, thus, would be interesting to apply the RCC method to electron-atom scattering problems and compare the results with the available 
experimental and other theoretical calculations.

On the other hand, study of atomic processes in plasma environment has received great attention in recent years due to its importance in 
understanding different physical processes involved in varieties of low to high temperature plasma (see a review \cite{plasma-rev} and references
therein). Most of such plasmas contain different charged species as well as free electrons which screen the atomic potential of the embedded atomic 
systems (neutral atoms or ions) as well as influence the collisional processes therein \cite{salzman,Weisheit,Murillo}. As a result, there are 
deviations in the structures of the atomic systems and associated collisional parameters from their corresponding behavior in the absence of plasma
environment \cite{Murillo,Saha,Sil}.  It is, therefore, important to understand the influence of the plasma environment on the atomic systems for plasma
diagnostic purposes. However, studying the electronic structures of atomic systems immersed in plasma environment is quite challenging and the 
accurate evaluation of the electronic structures of atoms or ions as well as adequate description of the atomic processes involved can be said to 
be one of the most demanding research in the recent years due to their wide range of applications. To determine atomic wave functions of the plasma 
embedded atomic systems, a suitable model is required which can account for the appropriate screening effects of the Coulomb potentials. 
Moreover, it is necessary to employ a many-body method that can include both the electron-electron correlations and relativistic effects 
rigorously. For this purpose various models have been proposed and used; viz., ion sphere model, Thomas Fermi model, Debye-H\"uckle model,
etc. \cite{Murillo,Thomas-Fermi,Ion-sphere-model,sil2}. For low density and high temperature plasma, the Debye-H\"uckel model is found to be
quite reliable to describe the influence of the screened Coulomb potentials in the calculations of atomic structure and collision 
parameters \cite{Murillo}. 

In any plasma the most dominant collision process that can occur is due to electron impact on atomic species causing excitation and ionization
processes. So far, there have been some efforts to understand the effects of plasma environment on the elastic and inelastic scattering of 
electrons from H-like and He-like ions \cite{Zammit-1,Zammit-2,Zhang-1,Zhang-2,Qi-1,Qi-2,Ghoshal}. However, most of such studies have been 
restricted to the total excitation cross-sections. It is only recently that Chen et al \cite{Chen} applied a fully relativistic distorted wave
theory to study the influence of Debye plasma on the magnetic cross-section and linear polarization of X-ray radiation following inner-shell
electron-impact excitation of highly charged Be-like ions. Therefore, there is in general lack of studies related to the
plasma screening effects on electron impact excitation of many-electron atomic systems and even so, for their magnetic sublevel cross-sections, 
and characteristic photon emission during the radiative de-excitation processes.  

There are two fold objectives here; first, to achieve accurate description of the electronic structure of many-electron atomic 
species and then, extending it to study the electron impact excitations of these atoms in the Debye plasma description so that collisional
process can also be described to the same degree of accuracy. The RCC method has been widely applied for high precision atomic structure 
calculations \cite{bks1,bks2,bks3,bks4}. In this work, we intend to employ this method to calculate the amplitudes of the electron excitations
of atomic systems. For illustration, we consider a singly charged magnesium ion (Mg$^+$) to study electron impact excitation of its $3s ~ ^2S_{1/2}
- 3p ~ ^2P_{1/2;3/2}$ resonant transitions in the plasma environment with the Debye-H\"uckel model framework. These resonance lines yield 
important emission features in the spectra of active galactic nuclei and provide information about intervening materials along the line of sight
\cite{Mg+-importance-1,Mg+-importance-2,Mg+-importance-3}. Since magnesium is one of the most abundant elements in the Galaxy, variation in the 
intensity of the spectral lines of the above transitions can act as diagnostic tools to determine the physical conditions of interstellar clouds. Thus,
for further understanding of the interstellar medium, it is important to study the electron impact excitations of the first excited state 
fine-structure splitting from the ground ($3s ~ ^2S_{1/2}$) state of the Mg$^+$ ion in the plasma environment. These transitions exist 
naturally in many astrophysical objects \cite{Mg-abundance-1,Mg-abundance-2}. So far, a few theoretical and experimental investigations have been devoted to understand
excitation processes in the plasma isolated Mg$^+$ ion \cite{LS-Mg+,Smith, William-1,William-5CC,YKKim,Pangantiwar}. However, such studies have
never been carried out for plasma embedded Mg$^+$. From this point of view, we perform calculations of differential cross-sections (DCS),
integrated cross-sections (ICS) and polarization of characteristic photon emission as function of Debye length considering various incident
electron energies ranging from 10 to 100 eV. To obtain these results accurately, it is imperative to evaluate both the bound and continuum wave functions
by including the Debye-H\"uckle interaction potentials accounting for the plasma screening effects at various plasma strengths.

In the point of view of above, we have implemented and used the RCC method in the singles and doubles excitation approximation (CCSD method) to obtain the scattering amplitudes of the electron from the Mg$^+$ ion at various plasma
strengths. The wave functions of the initial and final states of the Mg$^+$ ion are evaluated in the similar approach adopted to
study ionization potential depressions (IPD) of the plasma embedded atomic systems \cite{BKS-1,BKS-2}. Also, our previously reported RDW method \cite{LS-Mg+} is 
extended further to calculate the wave function of the scattered electron in the vicinity of the plasma embedded atomic systems. We 
consider the Debye-H\"uckel screening potential model and solve the coupled Dirac equations using distortion potential, which is taken as the 
spherically averaged static potential of the target ion.  

This paper is organized as follows: Method of calculations for the scattered electron and target wave functions in the plasma environment as
well as evaluation of scattering amplitude are mentioned briefly in Sec. \ref{sec2}. In Sec. \ref{sec3}, we present results and their 
discussions following the concluding remarks in Sec. \ref{sec4}. Unless stated otherwise, we adopt the atomic units (a.u.) throughout the paper.

%=========================================
%-----------------Theory------------------
%=========================================
%
\section{Theory and Methodology} \label{sec2}

\subsection{Theory of scattering amplitude}
Within the relativistic distorted wave approximation the first order direct scattering amplitude for
 electron impact excitation of an ion having nuclear charge $Z$ and number
of electrons $N$ from its initial state $| \Psi_i \rangle$ to final state $| \Psi_f \rangle$ can be expressed as
\begin{widetext}
\be
\label{genTmatrix}
f(J_i,M_i,\mu_i;\,J_f,M_f,\mu_f,\theta)= 4 \pi^2 \sqrt{\frac{k_f}{k_i}} \Big\langle \Psi_f(\textbf{1,2,..,N})\,
F_f^{DW-}(\textbf{k}_f,\textbf{N+1}) \left\vert V_{in} - V_d({r_{N+1}}) \right\vert
\\\nonumber
\times
 \Psi_i(\textbf{1,2,..,N})\, F_i^{DW+}(\textbf{k}_i,\textbf{N+1})\, \Big\rangle \,.
\ee
\end{widetext}

In this amplitude the ionic states are uniquely specified with well-defined total angular momentum $J$ and its magnetic component $M$,
whereas spin projections of the projectile and scattered electrons are denoted respectively, by $\mu_i$ and $\mu_f$.  $|\Psi_{i,f} (N) \rangle$
show the $N$-electron target wave functions having position coordinates with respect to the nucleus as $r_j$, where $j = 1,2,..,N$, and
$F^{DW+(-)}_{i,f}$ with position coordinate $r_{N+1}$ represent the incident or scattered electron and its superscript $DW$ means they 
are determined by the distorted wave function approximation. In our RDW method, they are calculated in the presence of a suitably defined
distortion potential $V_d$. Here $-$ ($+$) signs refer to the incoming (outgoing) waves with their relativistic wave numbers $k_{i}$ 
($k_f$). In Eq. (\ref{genTmatrix}), $V_{in}$ is the screened Coulomb interaction between the projectile electron and the target ion. The 
geometry of the scattering process is chosen so that the direction of the incident electron defines the $z$ axis and $xz$ plane forms the
scattering plane. Therefore, the scattering angle $\theta$ is the angle between the wave vectors $\textbf{k}_i$ and $\textbf{k}_f$ of 
the incident and scattered electrons.

In order to evaluate the scattering amplitude, Eq.~(\ref{genTmatrix}) can be casted into the following form
\begin{widetext}
\be
\label{scat-amp}
f(J_i,M_i,\mu_i;\,J_f,M_f,\mu_f,\theta)= 4 \pi^2 \sqrt{\frac{k_f}{k_i}} \Big\langle
F_f^{DW-}(\textbf{k}_f,\textbf{N+1}) \left\vert U_{i \rightarrow f}(r_{N+1}) \right\vert
  F_i^{DW+}(\textbf{k}_i,\textbf{N+1})\, \Big\rangle \,,
\ee
\end{widetext}
where
\be
\label{V_i-f}
U_{i \rightarrow f}(r_{N+1})=
\Big\langle \Psi_f(\textbf{1,2,...N}) \,
\left\vert V_{in}
\right\vert
\Psi_i(\textbf{1,2,...N}) \,
\Big\rangle \,.
\ee
In the following subsections we describe in detail various constituents that are required to evaluate the above scattering amplitude.

\subsection{Coulomb potentials}

The atomic electrons as well as free electrons will experience the screening of electron-nucleus as well as electron-electron Coulomb 
interactions. Therefore, $V_{in}$ can be written as
\be
\label{Debye-potential}
V_{in}(r_i,r_j)= V_{n}(r_i) + V_{ee}(r_{ij}),
\ee
 where $V_{n}(r_i)$ and $V_{ee}(r_{ij})$ represent, respectively, the interaction of bound or free $i^{th}$ electrons with atomic
 nucleus and the other $j^{th}$ electron. In the following we discuss the general forms, as described by Debye-H\"uckle model, for 
$V_{n}(r_i)$ and $V_{ee}(r_{ij})$ that are used in the present work to evaluate the scattering amplitude as well as in obtaining the 
target and projectile electron wave functions in the plasma. The same can also be deduced for the plasma free systems by taking 
special conditions.

In the weakly coupled plasma, an electron of the plasma embedded atomic system located at $r_i$ can see the screened nuclear potential
with Fermi nuclear charge distribution as
\begin{eqnarray}
 V_{n}(r_i) = -\frac{Z e^{-r_i/D_l}}{\mathcal{N}r_i} \times \ \ \ \ \ \ \ \ \ \ \ \ \ \ \ \ \ \ \ \ \ \ \ \ \ \ \ \ \ \ \ \ \ \ \ \ \ \ \ \  \nonumber\\
\left\{\begin{array}{rl}
\frac{1}{b}(\frac{3}{2}+\frac{a^2\pi^2}{2b^2}-\frac{r^2}{2b^2}+\frac{3a^2}{b^2}P_2^+\frac{6a^3}{b^2r}(S_3-P_3^+)) & \mbox{for $r_i \leq b$}\\
\frac{1}{r_i}(1+\frac{a62\pi^2}{b^2}-\frac{3a^2r}{b^3}P_2^-+\frac{6a^3}{b63}(S_3-P_3^-))                           & \mbox{for $r_i >b$} .
\end{array}\right.
\label{eq12}
\end{eqnarray}
Here $D_l$ refers to the Debye length expressed as
\be
\label{Debye-length}
D_l = \sqrt{\frac{k_B T_e}{4 \pi (1+z_i) n_e}},
\ee
where $k_B$ is the Boltzmann constant, $n_e$ is the electron density, $T_e$ is the temperature and $z_i$ is charge of the target in
the plasma. In Eq. (\ref{eq12}) the various factors are given by
\begin{eqnarray}
\mathcal{N} &=& 1+ \frac{a^2\pi^2}{b^2} + \frac{6a^3}{b^3}S_3  \nonumber \\
\text{with} \ \ \ \ S_k &=& \sum_{l=1}^{\infty} \frac{(-1)^{l-1}}{l^k}e^{-lb/a} \ \ \  \nonumber \\
\text{and} \ \ \ \ P_k^{\pm} &=& \sum_{l=1}^{\infty} \frac{(-1)^{l-1}}{l^k}e^{\pm l(r-b)/a} .
\end{eqnarray}
In the above expressions, $b$ is known as the half-charge radius and $a= 2.3/4(\ln3)$ is related to the skin thickness of the nucleus. The
parameter $b$ is evaluated using the relation
\begin{eqnarray}
b&=& \sqrt{\frac {5}{3} r_{rms}^2 - \frac {7}{3} a^2 \pi^2}\,,
\end{eqnarray}
with the appropriate value of the root mean square radius of the nucleus $r_{rms}$, which is estimated using the empirical formula
\begin{eqnarray}
 r_{rms} =0.836 A^{1/3} + 0.570 \,,
\end{eqnarray}
in fm for the atomic mass $A$.

The screened two-body Coulomb potential within the atomic system can be expressed as
\begin{eqnarray}
V_{ee}(r_{ij}) &=& \sum_{j \ge i}^{N}\frac{e^{- |r_{i}-r_j|/D_l} }{|r_{i}-r_j|}  \nonumber \\
       &=& \frac{4\pi}{\sqrt{r_ir_j}}\sum_{k=0}^{\infty}I_{k+\frac{1}{2}}( r_< / D_l) K_{k+\frac{1}{2}} ( r_> /D_l ) \nonumber \\
       && \times \sum_{q=-k}^k Y_{q}^{k\ast}(\theta,\phi)Y^k_{q}(\theta,\phi) ,
\label{veff}
\end{eqnarray}
where $I_{k+\frac{1}{2}}(r)$ and $K_{k+\frac{1}{2}}(r)$ are the modified Bessel functions of the first and second kind, respectively, with $r_> =$
max($r_i,r_j$); $r_<=$ min($r_i,r_j$), and $Y^k_{q}(\theta,\phi)$ is the spherical harmonics of rank $k$ with its component $q$. In terms of
the Racah operator ($C^k_{q}$), the above expression is given by a scalar product as
\begin{eqnarray}
V_{ee}(r_{ij}) &=& \frac{1}{\sqrt{r_ir_j}}\sum_{k=0}^{\infty} (2k+1) I_{k+\frac{1}{2}} ( r_< /D_l) K_{k+\frac{1}{2}} (r_> / D_l) \nonumber \\ && \times \mbox{\boldmath$\rm C$}^k(\hat{r}_i) \cdot \mbox{\boldmath$\rm C$}^k(\hat{r}_j).
\end{eqnarray}

Similarly, the screened two-body Coulomb potential for interaction of projectile electron with atomic electrons can be expressed as
\begin{eqnarray}
V_{ee}(r_{j,N+1}) &=& \sum_{j =1}^{N}\frac{e^{- |r_j-r_{N+1}|/D_l} }{|r_j-r_{N+1}|} .
\end{eqnarray}
This expression can further be written as multipole expansion in the same way as done above in Eq.~(\ref{veff}).

It to be noted that in the plasma free system, $n_e=0$, so $D_l \rightarrow \infty$. Considering this special condition, results for the 
plasma isolated systems are obtained.

\subsection{RCC method for the wave functions of the target}

In the relativistic framework, the wave functions of the target is determined by considering the Dirac-Coulomb (DC) Hamiltonian, which 
is given by
\begin{eqnarray}
H &=& \sum_{i=1}^{N} \left [ c\mbox{\boldmath$\alpha$}_i\cdot \textbf{p}_i+(\beta_i -1)c^2 + V_{n}(r_{i})
  + \sum_{j\ge i} V_{ee}(r_{ij}) \right ], \nonumber
  \label{NSDeq}
\end{eqnarray}
where $\mbox{\boldmath$\alpha$}$ and $\beta$ are the Dirac matrices and $c$ is the velocity of light.

The initial and final states of the considered target Mg$^+$ has been calculated by expressing them as a common inert core, i.e.
$[2p^6]$, and a valence electron in the respective states. Conveniently, we obtain first the single particle orbitals of the closed-core
employing the Dirac-Hartree-Fock (DHF) method. The wave functions of the exact states are evaluated by using the exponential
{\it ansatz} in the RCC theory as
\begin{eqnarray}
 \vert \Psi_v \rangle  = e^T \{ 1+ S_v \} \vert \Phi_v \rangle ,
 \label{eqcc}
\end{eqnarray}
where $\vert \Phi_v \rangle=a_v^{\dagger} \vert \Phi_0 \rangle$, for the DHF wave function of the closed-core $|\Phi_0 \rangle$. Here,
$T$ and $S_v$ are the RCC excitation operators that excite electrons from $\vert \Phi_0 \rangle$ and $\vert \Phi_v \rangle$, respectively,
to the virtual space. It can be noted that the above expression is linear in $S_v$ operator owing to presence of only one valence orbital
$v$ in $\vert \Phi_v \rangle$. This expression is, however, exact and it accounts for the non-linear effects through the products of $T$ and $S_v$ operators. In
the CCSD method approximation, the RCC operators are expressed as $T=T_1+T_2$ and $S_v=S_{1v}+S_{2v}$ with the subscripts 1 and 2
referring to the singly and doubly excited state configurations, respectively. The amplitudes of these RCC operators are evaluated
by solving the following coupled-equations
\begin{eqnarray}
 \langle \Phi_0^* \vert \overline{H}  \vert \Phi_0 \rangle &=& 0
\label{eqt}
 \end{eqnarray}
and
\begin{eqnarray}
 \langle \Phi_v^* \vert \big ( \overline{H} - \Delta E_v \big ) S_v \vert \Phi_v \rangle &=&  - \langle \Phi_v^* \vert \overline{H}_N \vert \Phi_v \rangle ,
\label{eqsv}
 \end{eqnarray}
where $\vert \Phi_0^* \rangle$ and $\vert \Phi_v^* \rangle$ are the excited state configurations, here up to doubles, with
respect to the DHF wave functions $\vert \Phi_0 \rangle$ and $\vert \Phi_v \rangle$ respectively. Here $\overline{H}= \big ( H e^T
\big )_c$ with subscript $c$ represents for the connected terms only and $\Delta E_v$ is the electron attachment energy (EA), which
is equivalent to second ionization potential (IP), of the electron of the valence orbital $v$. We evaluate $\Delta E_v$ by
\begin{eqnarray}
 \Delta E_v  = \langle \Phi_v \vert \overline{H} \left \{ 1+S_v \right \} \vert \Phi_v \rangle - \langle \Phi_0 | \overline{H} | \Phi_0 \rangle .
 \label{eqeng}
\end{eqnarray}
Both Eqs. (\ref{eqsv}) and (\ref{eqeng}) are solved simultaneously, as a result Eq. (\ref{eqsv}) effectively becomes non-linear in
the $S_v$ operator. In fact, the excitation energy (EE) between two given states is evaluated by taking difference between the
respective EAs obtained from the above procedure.

After obtaining amplitudes of the RCC operators using the above described equations, the matrix element of $V_{in}$ between the states
$\vert \Psi_i \rangle$ and $\vert \Psi_f \rangle$, i.e. Eq.~(\ref{V_i-f}), is evaluated using the expression
\begin{eqnarray}
U_{i \rightarrow f}(r_{N+1}) &=&
\langle \Psi_f| V_{in}| \Psi_i\rangle
\nonumber \\
&=& \frac{\langle\Phi_f|\tilde{V}_{in}|\Phi_i\rangle}{\sqrt{\langle\Phi_f|\{1+\tilde{N}_f\}|\Phi_f\rangle
\langle\Phi_i|\{1+\tilde{N}_i\}|\Phi_i \rangle}} , \nonumber \\
\label{eqno}
\end{eqnarray}
where $\tilde{V}_{in}=\{1+S_f^{\dagger} \} e^{T^{\dagger}} V_{in} e^T \{1+S_{i}\}$ and $\tilde{N}_{k=f,i}=\{1+S_k^{\dagger} \}
e^{T^{\dagger}}e^T \{1+S_{k}\}$. As can be seen, it involves two non-terminating series in the numerator and denominator in the above
expression, which are $e^{T^{\dagger}} V_{in} e^T$ and $e^{T^{\dagger}} e^T$ respectively. We adopt iterative procedures to account for
contributions from these non-terminating series as have been described in our earlier works on the plasma isolated systems \cite{bks3,bks4}.

\begin{table}[t]
\caption{EAs (in eV) and EEs (in eV) as function of $D_l$ (in a.u.) in the plasma embedded Mg$^+$ ion. Results for the plasma free ($D_l \rightarrow \infty$)
are compared with the NIST data \cite{NIST}.}
\begin{center}
\begin{tabular}{c c c c c c c}\hline\hline
\multirow{3}{*}{ $D_l$} & \multicolumn{4}{c}{EE} & & \\
                & \multicolumn{2}{c}{$3s~^2S_{1/2}-3p~^2P_{1/2}$} & \multicolumn{2}{c}{$3s~^2S_{1/2}-3p~^2P_{3/2}$} & \multicolumn{2}{c}{EA} \\
\cline{2-3} \cline{4-5} \\
        & RCC        &   NIST  &     RCC        &   NIST  &     RCC        &   NIST \\\hline
         $\infty$   &  4.4162   & 4.4224    &    4.4283      & 4.4348  &    7.6909       &7.6460    \\
         100        &  4.4142   &     -     &    4.4263      & -      &    7.4208       &-    \\
         10         &  4.2489   &   -       &    4.2599      & -      &    5.1766       &-    \\
         7.0        &  4.0908   &    -      &    4.1007      & -      &    4.2408       &-    \\
         5.75       &  3.9472   &    -      &    3.9562      & -      &    3.6205       &-    \\
         3.80       &  3.3953   &    -      &    3.4012      & -      &    2.1173       &-    \\
         2.865      &  2.6089   &   -       &    2.6111      & -      &    1.0679       &-    \\
         2.863      &  2.6061   &   -       &      -         & -      &    1.0655       &-     \\\hline\hline
\end{tabular}
\end{center}
\label{tab1}
\end{table}

\subsection{RDW method for the wave functions of the scattered electron}
The distortion potential $V_d$ is taken to be the spherically averaged static potential
of the ion in its initial state. It can be expressed as
\be
\label{V_d}
V_d=
\Big\langle \Psi_i(\textbf{1,2,...N}) \,
\left\vert V_{n}(r_{N+1}) + \sum_{j=1}^{N}V_{ee}(r_{j,N+1})
\right\vert
\\ \nonumber
\Psi_i(\textbf{1,2,...N}) \,
\Big\rangle \,.
\ee
After substituting $V_{ee}$  from Eq.~({\ref{veff}}) with $k=0$, $V_d$ is given by

\be
\label{V_d}
V_d(r_{N+1}) = V_{n}(r_{N+1}) + \pi\sum_{j=1}^{N}\int \frac{1}{\sqrt{r_j r_{N+1}}}
\sqrt{\frac{r_<}{r_>}}
\\\nonumber
\times
\sinh(r_</D_l)
 e^{-r_>/D_l}
 |\Psi_i|^2 dr_j dr_{N+1}.
\ee
It is assumed that the atomic wave functions are  not distorted by the projectile electron. The relativistic partial wave expansion, as
described in our previous work \cite{LS-Mg+}, has been used for distorted wave functions $F^{DW+(-)}_{i,f}$ of the projectile electron.
For each partial wave, the large and small components of the continuum wave functions are obtained by solving the coupled Dirac
equations using $V_d$ potential \cite{LS-Mg+}. Thus, scattering amplitude given by Eq. (\ref{scat-amp}) can finally be evaluated using 
both the partial wave expansion and the matrix element given by Eq.~(\ref{eqno}).

\begin{figure*}[t]
\includegraphics[trim = {8cm 0.3cm 8cm 2cm},scale=1.0]{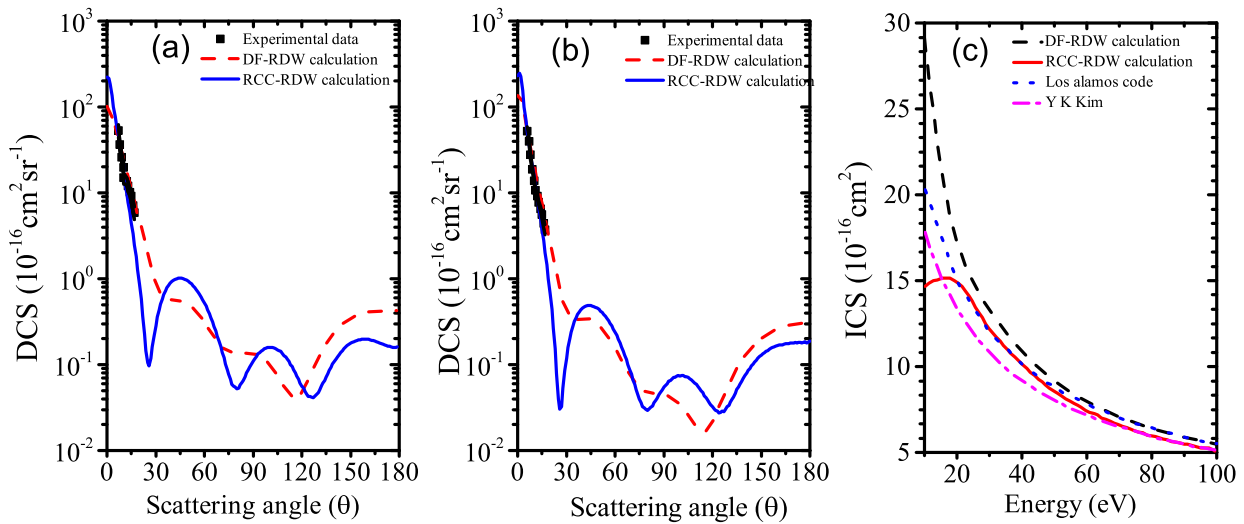}
\caption{(Color online) Electron impact excitations of the $3s~^2S_{1/2} \rightarrow 3p~^2P_{1/2}$ and $3s~^2S_{1/2} \rightarrow 3p~^2P_{3/2}$ 
transitions of Mg$^+$ ion. We show (a) DCS at 35 eV (b) DCS at 50 eV and (c) ICS with solid curve representing the RCC-RDW 
calculations, dashed curves for the DF-RDW calculations \cite{LS-Mg+}, short dashed curves for the calculations with the Los Alamos 
code \cite{LosAlamos}, and dashed-dotted curves for the scaled results from Ref. \cite{YKKim}. The available experimental values are shown
by points with error bars that were reported in Refs. \cite{William-1,William-5CC}.}
\label{fig1}
\end{figure*}

\subsection{Scattering Parameters}
After obtaining the scattering amplitude, the relevant parameters related to the electron-atom collision process can be calculated. With 
our normalization of the distorted waves, the DCS for excitation of ion from initial state $i$ to final state $f$ is given by
\be
\label{dcs}
\frac{d\sigma(\theta)}{d\Omega}=
\frac{1}{2(2J_i+1)}\sum_{\substack {M_i,\mu_i\\M_f,\mu_f}}
|f(J_i,M_i,\mu_i;\,J_f,M_f,\mu_f,\theta)|^2 \, .
\nonumber \\
\ee
In the above expression, we have summed over the spins of the incident and scattered electron as well as magnetic quantum numbers corresponding to
the total angular momenta of the ion in the initial and final states. The ICS (denoted by $\sigma$) is obtained by integrating the DCS over all
the scattering angles. However, if averaging is done over initial magnetic sublevels $M_i$, cross-section for excitation to a specific
final magnetic sublevel $M_f$ can be calculated (denoted by $\sigma_{M_f}$). Using the magnetic sublevel cross-sections, the degree of
linear polarization of the photon emission can be obtained. Here we consider excitations from the $3s ~ ^2S_{1/2}$ ground state to the
$3p ~ ^2P_{1/2}$ and $3p ~ ^2P_{3/2}$ levels and subsequent decay from the $3p ~ ^2P_{3/2}$ level to  the ground state by emitting a
photon. Since in the electron--photon coincidence experiments the photon is detected in the direction perpendicular to the scattering
plane,  the linear polarization of the emitted photon can be obtained using the $\sigma_{M_f}$ values of the excited state, i.e.,
\be
\label{P_L}
P=\frac{3(\sigma_{1/2}-\sigma_{3/2})}{3 \sigma_{3/2}+5 \sigma_{1/2}} \, .
\ee
The linear polarization of the photon emitted due to decay from the $3p ~ ^2P_{1/2}$ level to the $3s ~ ^2S_{1/2}$ state
will be zero, since cross-sections for the magnetic sublevels $M_f=\pm1/2$ are equal.

\begin{figure*}
\begin{center}
\begin{tabular}{cc}
\resizebox{70mm}{!}{\includegraphics[trim = 1cm 7cm 14cm 2cm,scale=0.5]{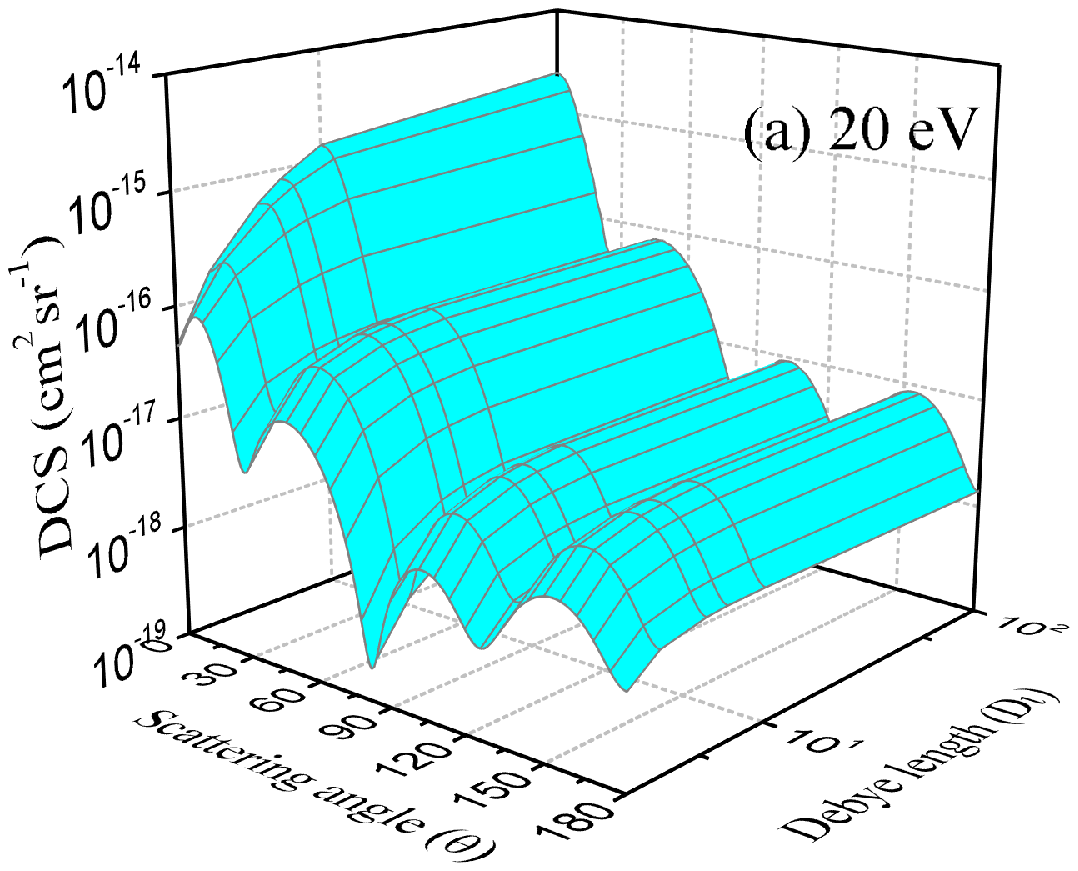}} &
\resizebox{70mm}{!}{\includegraphics[trim = 1cm 7cm 14cm 2cm,scale=0.5]{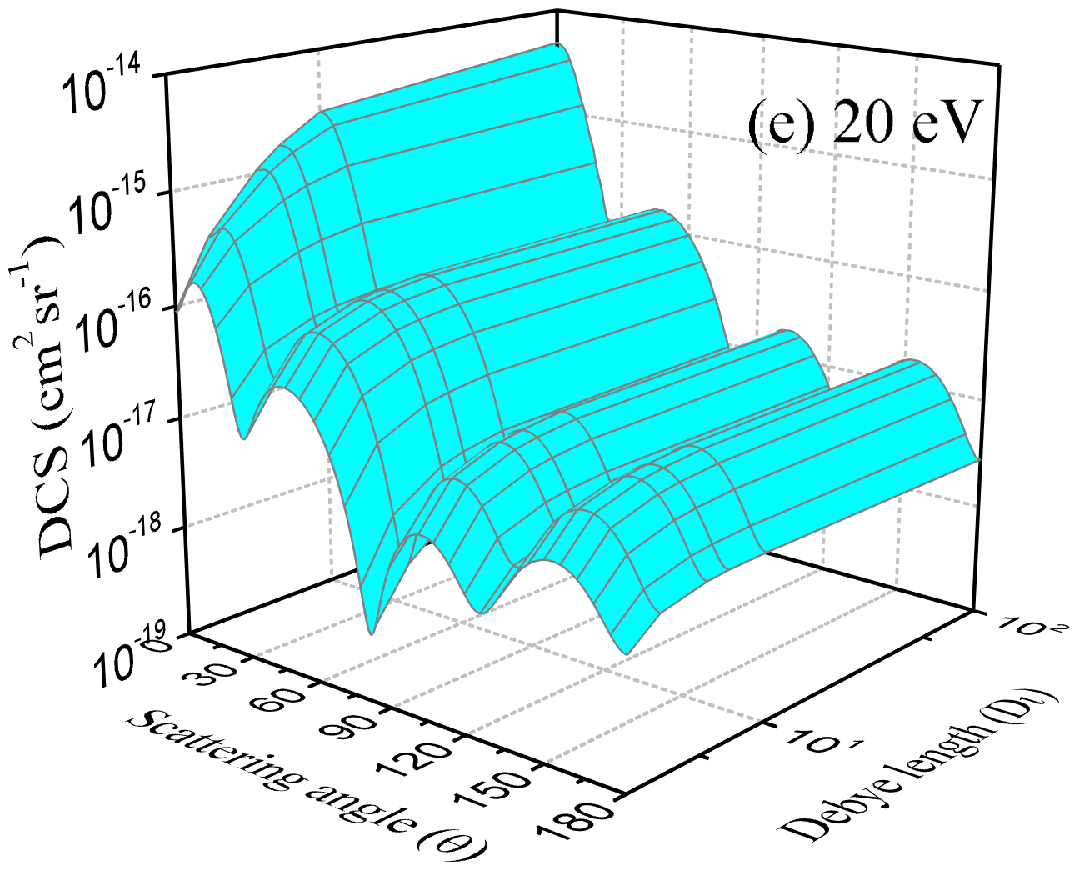}} \\
\resizebox{70mm}{!}{\includegraphics[trim = 1cm 7cm 14cm 4cm,scale=0.5]{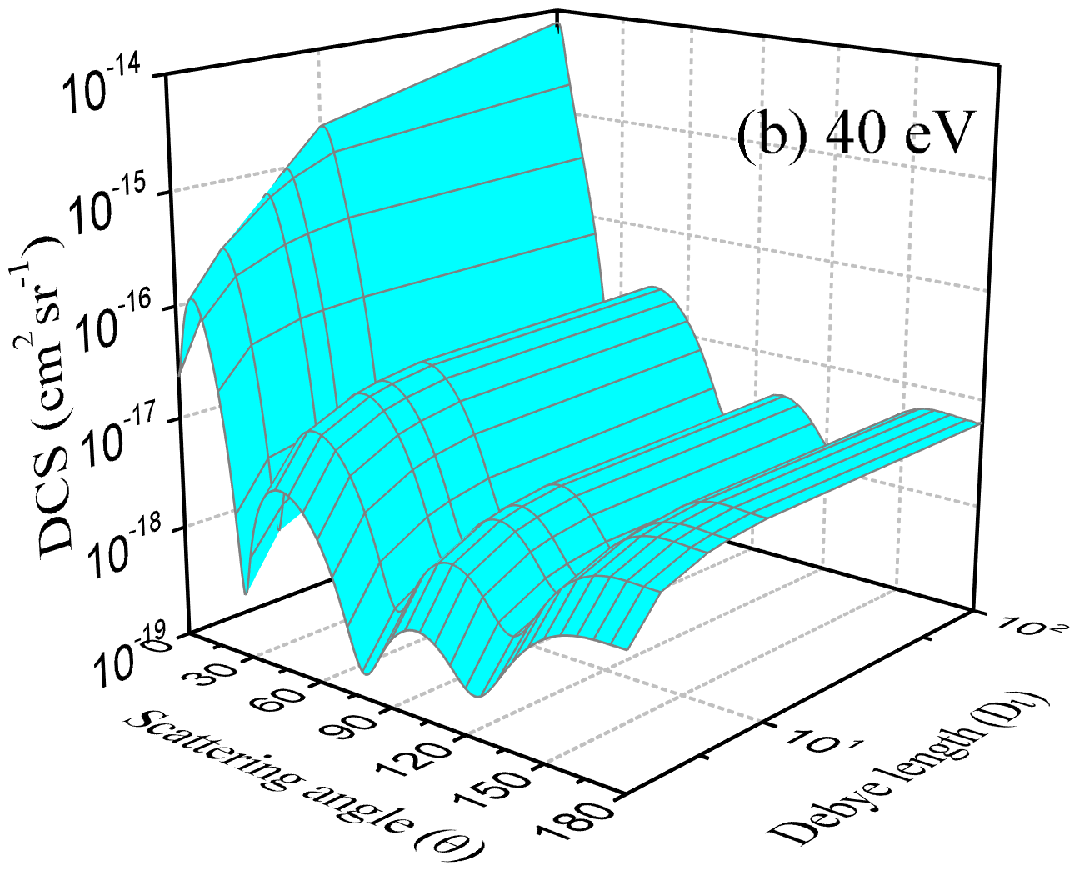}} &
\resizebox{70mm}{!}{\includegraphics[trim = 1cm 7cm 14cm 4cm,scale=0.5]{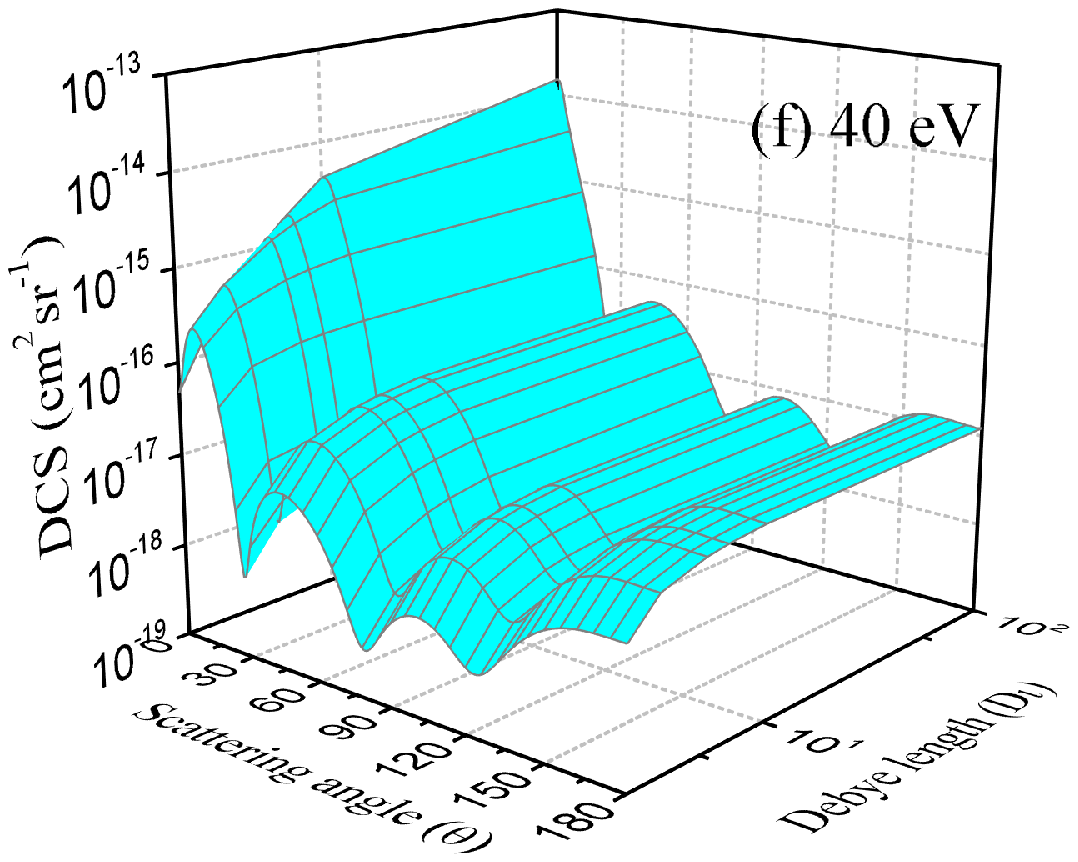}} \\
\resizebox{70mm}{!}{\includegraphics[trim = 1cm 7cm 14cm 4cm,scale=0.5]{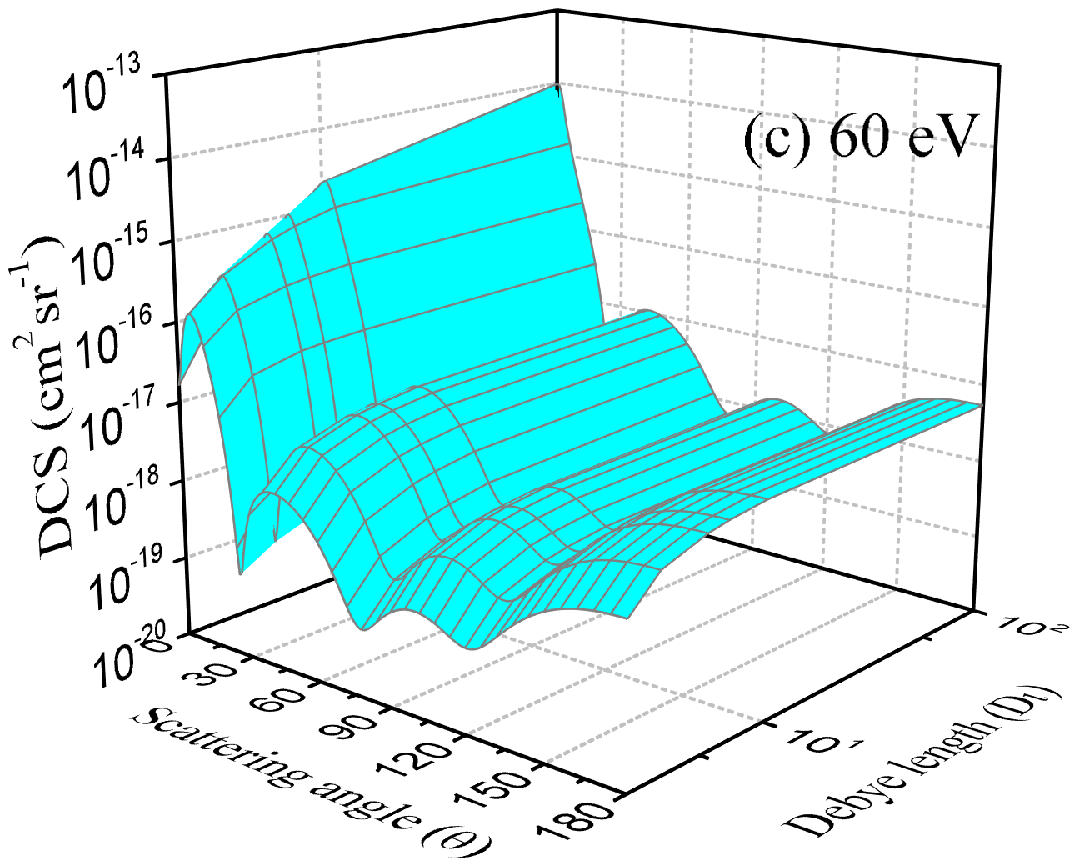}} &
\resizebox{70mm}{!}{\includegraphics[trim = 1cm 7cm 14cm 4cm,scale=0.5]{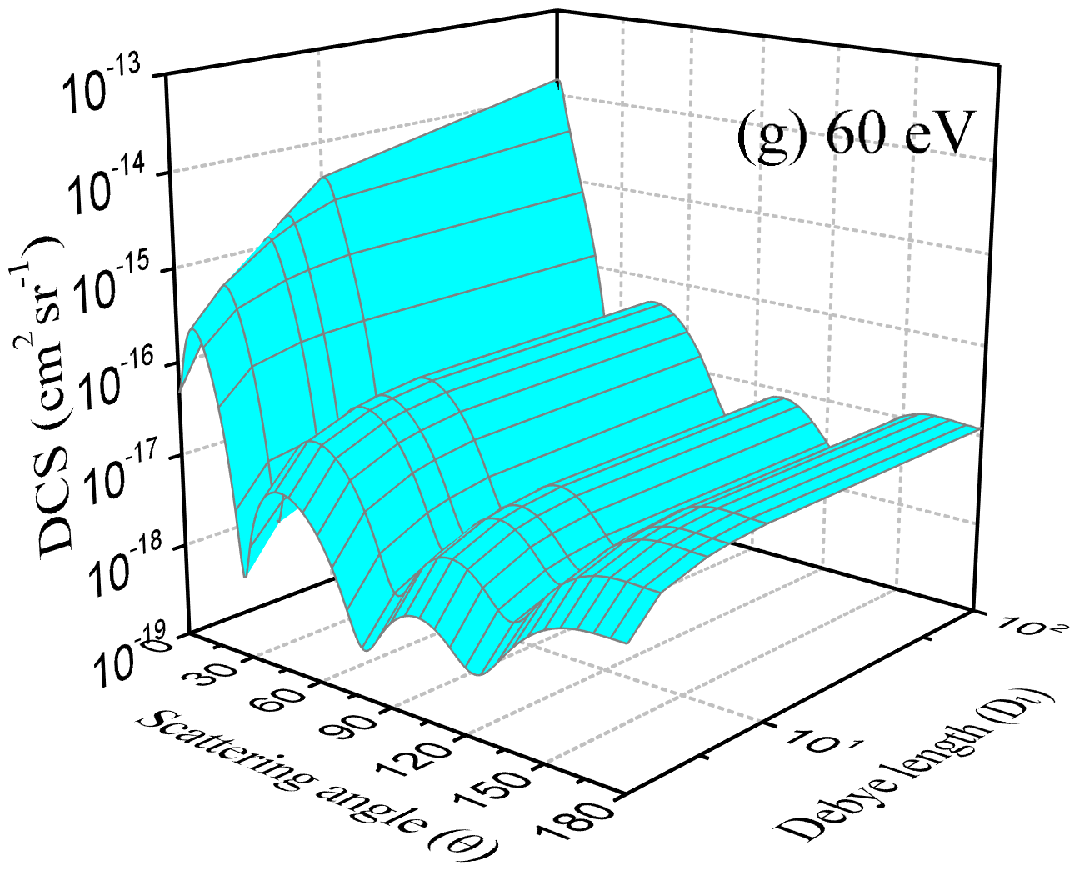}} \\
\resizebox{70mm}{!}{\includegraphics[trim = 1cm 10cm 14cm 4cm,scale=0.5]{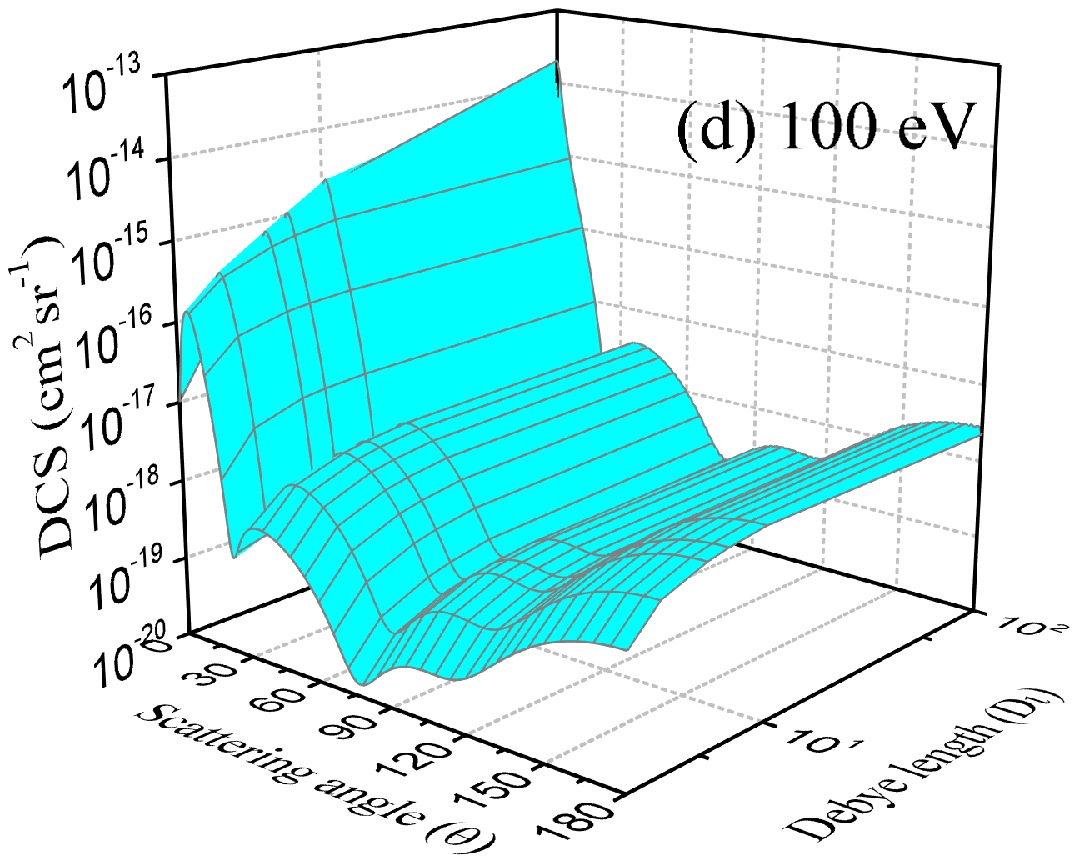}} &
\resizebox{70mm}{!}{\includegraphics[trim = 1cm 10cm 14cm 4cm,scale=0.5]{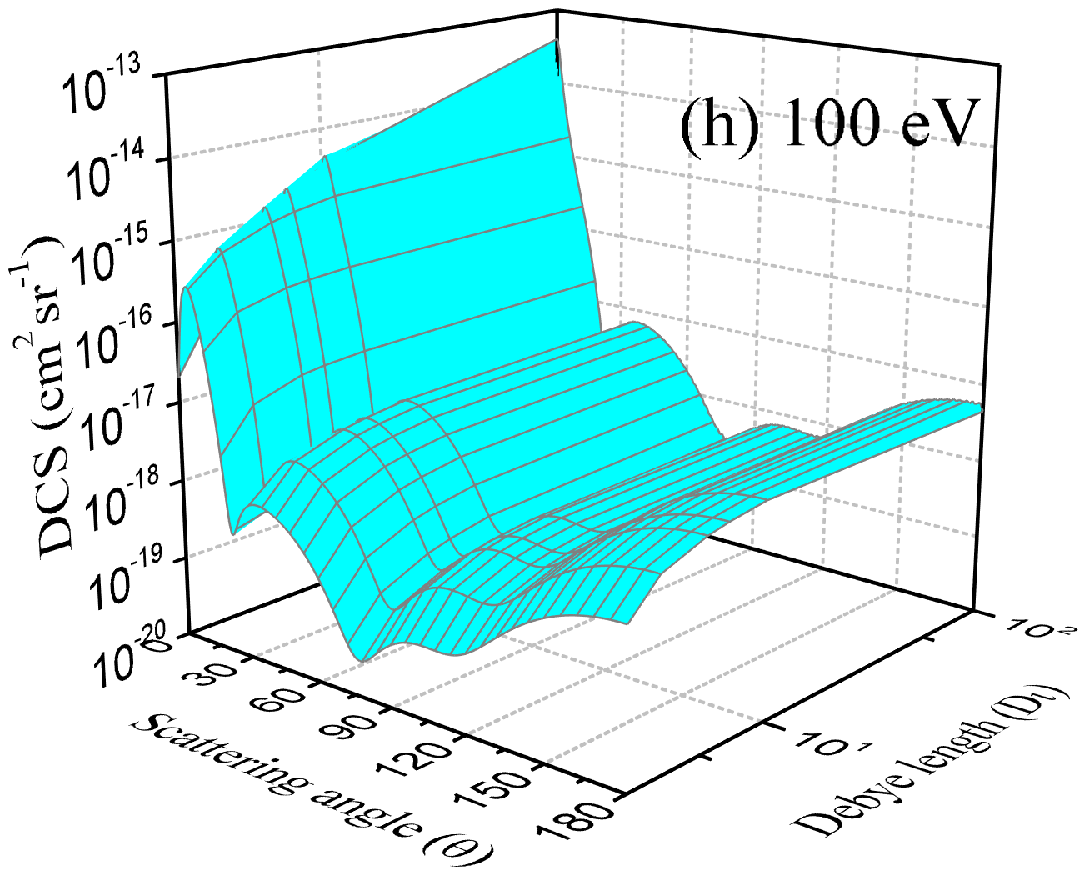}}
\end{tabular}
\caption{(Color online) Demonstration of variations in DCSs at various Debye lengths as function of incident electron energy
for the excitations from the 3$^2S_{1/2}$ state to the 3$^2P_{1/2}$ and 3$^2P_{3/2}$ states that are shown in the 
left and right panels, respectively.}
\label{fig2}
\end{center}
\end{figure*}

\section{Results and discussion}\label{sec3}

We present here our relativistic calculations for electron impact excitations of the
$3s ~ ^2S_{1/2}$ to $3p ~ ^2P_{1/2}$ and $3p ~ ^2P_{3/2}$ states in Mg$^+$.
We have obtained the DCS, ICS and linear polarization results
for isolated as well as plasma embedded Mg$^+$. It is evident that the reliability
of the collision parameters depends crucially on the accuracies of the
wave functions used in the calculations. Therefore, in Table I we have
compared the RCC results for EAs and EEs for the considered states in the present work with the corresponding experimental
values quoted in the NIST database \cite{NIST}. We find that our results are in excellent agreement with the NIST data.
We have also calculated these quantities at various $D_l$ values for which no other theoretical or experimental results are available
to compare. These $D_l$ values can correspond to different plasma conditions with varying temperature or varying electron density
that are of general interest. We notice that our EAs and EEs decrease with decreasing value of $D_l$. This feature is in confirmation
with earlier studies revealing IPD of the atomic systems in plasma environment \cite{BKS-1}. It can be further be seen from the Table \ref{tab1} that there is no result presented at $D_l$ = 2.863 a.u. for excitation energy of
the 3$p~ ^2P_{3/2}$ state. This is due to the reason that a slight change in the value of $D_l$ from 2.865 to 2.863 leads to the
conversion of the bound 3$p ~ ^2P_{3/2}$ state into a continuum state.

\begin{figure*}[t]
\includegraphics[trim = {8cm 0.3cm 8cm 2cm}, scale=1.0]{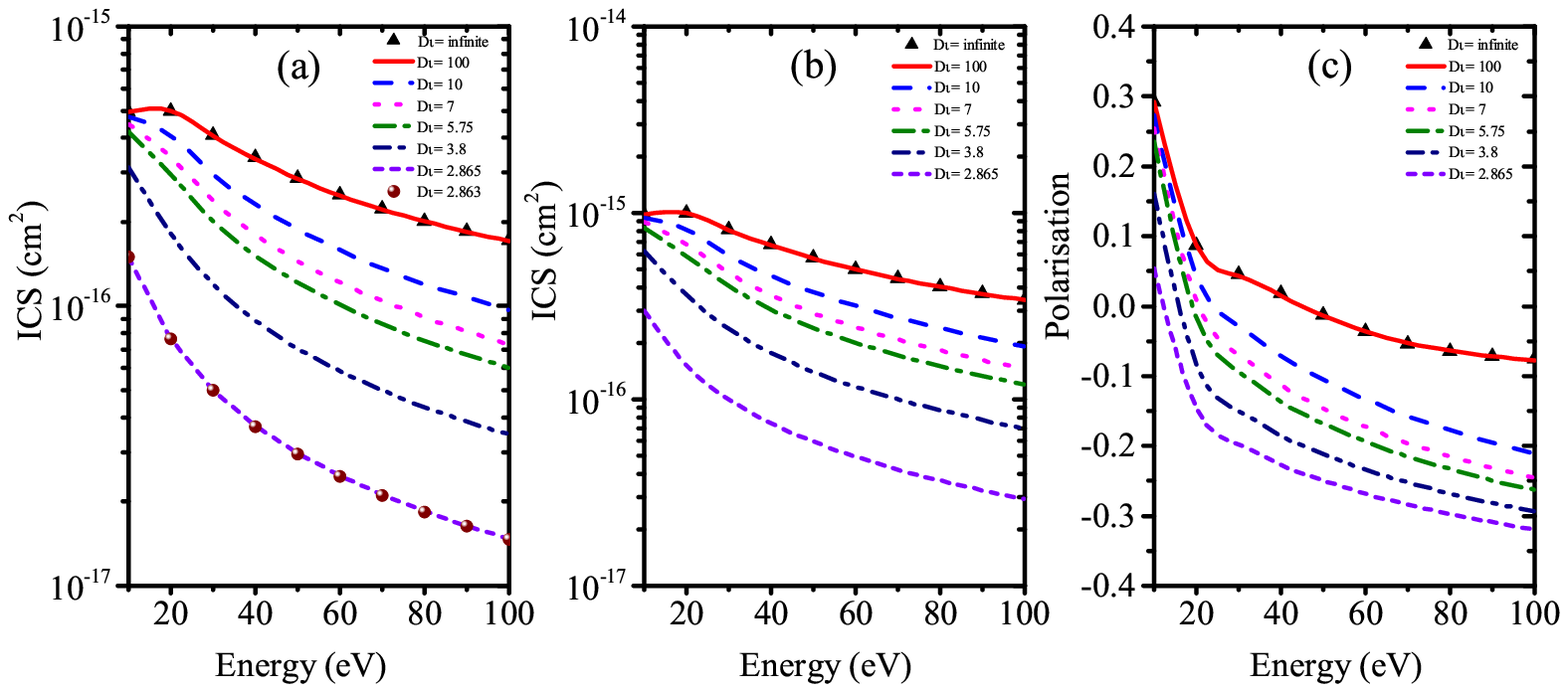}
\caption{(Color online) Demonstration of ICSs for the electron impact excitations in the plasma embedded Mg$^+$ ion. Results 
shown in (a) for the $3s~^2S_{1/2} \rightarrow 3p~^2P_{1/2}$ transition and shown in (b) for the $3s~^2S_{1/2} \rightarrow 3p~^2P_{3/2}$
transition. Results shown in (c) are the polarization fractions for the $3p~^2P_{3/2}$ to $3s~^2S_{1/2}$ transition.}
\label{fig3}
\end{figure*}

As mentioned already in Introduction, the RCC method is employed for the first time to study the
collisional excitation of atoms and hence, for understanding the influence of plasma environment on collisional
parameters. Therefore, it is worth to compare the new cross-section results, which we denote as RCC-RDW, using RCC wave functions
with the previously available theoretical and experimental results. To this effect, in Figs. \ref{fig1}(a) and (b) our DCS results
are shown, respectively, at 35 eV and 50 eV and are compared with the experimental data of Williams et al \cite{William-1,William-5CC}
and previous RDW calculations \cite{LS-Mg+}.  Since the measurements \cite{William-1,William-5CC} have been reported for the
sum of the cross-sections for the transition from 3$s ~^2S_{1/2}$ to 3$p ~ ^2P_{3/2}$ and 3$p ~ ^2P_{1/2}$ states, we have
presented the summed DCSs in Figs. \ref{fig1}(a) and (b).  For the sake of clarity we have not included  old theoretical results from 
non-relativistic distorted wave  \cite{Pangantiwar} and 5 state close-coupling \cite{William-5CC} methods. We have already reported
detailed comparison of  RDW results with these non-relativistic calculations and measurements in our previous work for singly
charged metal ions including Mg$^+$ \cite{LS-Mg+}. The difference between present RCC-RDW and old RDW calculations \cite{LS-Mg+}
 is due to the inclusion of higher correlation effects in our RCC method that is used in the present calculations to obtain bound state
wave functions of the Mg$^+$ ion. Earlier, these wave functions were obtained using the MCDF method from the readily available 
GRASP92 code \cite{grasp92}. Therefore, we refer to these results here as DF-RDW calculations. It can be seen from Figs. \ref{fig1}(a)
and (b) that our RCC-RDW results are in good agreement with the DF-RDW calculations and the measurements, which are available up to 
20$^0$ scattering angles. The disagreement between the shape of the theoretical DCS curves is more after 20$^0$ scattering angles. It
could be due to accounting electron correlation effects in the RCC method, especially the core-polarization effects, to all orders 
in perturbation. The shortcoming of the MCDF method is that it cannot incorporate core-polarization effects rigorously. Therefore, 
we believe that the present RCC-RDW results are improved ones that need to be investigated by the future experiments. This demands
for more measurements at large scattering angles to confirm the shapes of the DCS curves. In Fig. \ref{fig1}(c), we have compared our 
summed ICS results with the DF-RDW calculations, scaled cross-sections of Kim \cite{YKKim} as well as semi-relativistic distorted wave 
calculations from the Los Alamos code \cite{LosAlamos}. The agreement of the new ICS curve is better with Kim's results in the entire 
energy range and it improves with increasing incident energy of the electron.

In Fig. \ref{fig2}, we have shown the DCS results at incident electron energy ranging from 20 to 100 eV and various Debye lengths viz.,
$D_l$ = 100, 10, 7, 5.75, 3.8, 2.865 and 2.863 a.u..  The left and right panels exhibit DCS results, respectively,  for the 3$^2S_{1/2}$ 
to 3$^2P_{1/2}$ and 3$^2S_{1/2}$ to 3$^2P_{3/2}$ transitions. Since 3$^2P_{3/2}$ state does not remain a bound state for $D_l < 2.865$ a.u.,
there is no DCS displayed for excitation of this state at $D_l < 2.863$ a.u.. Further, we find that at $D_l = 100$ a.u., the DCS
curves overlap with those at $D_l = \infty$, which corresponds to no plasma screening, at all the projectile electron energies.
As the value of $D_l$ decreases, the DCS decreases consistently while maintaining the similar shape of the curves approximately beyond
10$^0$ scattering angles. This can be attributed to the fact that the screening of nuclear charge
increases with decreasing Debye length, causing a drop in the value of the cross-sections. The maxima of the DCS curves fall
by two orders of magnitude in the considered range of Debye length. Moreover, in the forward scattering angle region up to 10$^o$,
the shapes of the DCS curves for $D_l \le 10$ a.u. is significantly different from that at $D_l = 100$ a.u..
Particularly, at incident electron energies greater than 40 eV, the DCS at $D_l = 100$ a.u. shows a sharp forward peak
which is characteristic of allowed transition, whereas, DCS at $D_l \le 10$ a.u. exhibit a broad peak at around 5$^0$, a well known
feature of dipole forbidden transitions.

The ICS for excitation of the ground state 3$^2S_{1/2}$ to the excited states 3$^2P_{1/2}$ and 3$^2P_{3/2}$ are displayed, respectively,
in Figs. \ref{fig3}(a) and (b). These results are shown for isolated ion as well as for various Debye lengths
$D_l$ ranging from 2.863 to 100 a.u. for $^2P_{1/2}$ state and from 2.865 to 100 a.u. for the 3$p~^2P_{3/2}$ state. The ICSs at 
$D_l=$ 100 a.u. are close to those for plasma-free ion. As observed for the DCS curves, the values of ICS also decrease with increasing
 plasma strength i.e., decrease in the value of $D_l$. The effect of plasma screening on the degree of linear polarization $P$ of the
 characteristic photon emitted due to decay of the $3p ~ ^2P_{3/2}$ state to the ground $3s ~ ^2S_{1/2}$ state is presented in Fig.
 \ref{fig3}(c). It can be observed that at 10 eV as $D_l$ decreases from 100 to 2.865 a.u.,  the value of $P$ decreases from 
 29\% to 5\%. In the isolated Mg$^+$, the polarization curve crosses zero at about 45 eV. However, in case of plasma embedded Mg$^+$ it 
 passes through zero at smaller values of energies with increasing plasma strength. This indicates that the magnetic sub-state with 
 $M_f=3/2$ gets more populated as compared to that with $M_f=1/2$ in the plasma environment. Moreover, the absolute value of $P$ 
 increases significantly showing enhancement of polarization to a value close to 32\% at 100 eV.

 \section{Conclusion}\label{sec4}
We have applied the combined RCC theory and RDW method to study the electron impact excitations of the resonant transitions, i.e. 
$3s ~ ^2S_{1/2}$ to $3p ~ ^2P_{1/2}$ and $3p ~ ^2P_{3/2}$ transitions, in the Mg$^+$ ion.  Excellent agreement of the 
attachment and excitation energies with the corresponding experimental values confirm on the accuracies of the wave functions of the 
ion obtained using the RCC theory. Observation in the depression of the ionization potential in the Mg$^+$ ion indicates about the 
effects of the plasma environment on the atomic structure. Very good agreements between the present DCS results at 35 and 50 eV  
with the available measurements and DF-RDW calculations in the plasma isolated Mg$^+$ ion indicate on the reliability of our 
calculations. Observed differences between the RCC-RDW and DF-RDW results at large scattering angles demand to ascertain them 
by carrying out further measurements. This can test validity of the many-body methods. Nevertheless, reasonable agreements between 
our theoretical results with the available experimental data suggest that results obtained in the plasma embedded ion are of similar 
accuracies. Results for the plasma embedded ion are reported as function of Debye length $D_l$ for the DCS, ICS and linear polarization
of the photon emission after decay of the $3p ~ ^2P_{3/2}$ state to the ground state at the incident electron energy ranging from 10 to
100 eV. It is observed from our analysis that presence of plasma environment affect the shape of the DCS curves in the small scattering
angle range such that the maxima of the DCS curves shifts from 0$^0$ towards 10$^0$ with decreasing Debye length. It is also found that 
the cross-sections decrease with decreasing value of $D_l$. There is enhancement in the degree of linear polarization of the photon
emission as the incident electron energy and plasma screening strength increase. These data could be useful for the plasma diagnostic 
processes in the astrophysical and laboratory plasmas.

\section*{Acknowledgements}

B.~K.~S. thanks Dr. Madhulita Das for many useful discussions and acknowledges financial support from CAS through the PIFI fellowship
under the project number 2017VMB0023. L.~S. and R.~S. acknowledge the support from the Department of Science and Technology (DST),
New Delhi for this work. P.~M. is thankful to Ministry of Human Resource and Development (MHRD) for providing her research assistantship.

%=================References===================
%


\begin{thebibliography}{20}

\bibitem{anil}
A. K. Pradhan and S. N. Nahar, {\it Atomic Astrophysics and Spectroscopy}, Cambridge University Press, New York (2011). 

\bibitem{johnson}
R. E. Johnson, {\it Introduction to Atomic and Molecular Collisions}, Plenum Press, New York and London, (1982).

\bibitem{amusia}
M. Y. Amusia, {\it Many-body effects in single photoionization processes}, Many-body atomic physics, Chapter 8, pg. 185, edited by J. J. Boyle and M. S. Pindzola, Cambridge University 
Press, New York (1998).

\bibitem{Post}
D. E. Post, J. Nucl. Mat. {\bf 220}, 143 (1995).

\bibitem{RDressler}
R. A. Dressler, Yu-hui Chiu, O. Zatsarinny, K. Bartschat, R. Srivastava, and L. Sharma, J. Phys. D \textbf{42},  185203 (2009).

\bibitem{Dipti}
Dipti, R. K. Gangwar, R. Srivastava and A. D. Stauffer, Eur. J. Phys. D \textbf{67},  40244 (2013).

\bibitem{Badnell}
N. R. Badnell, G. Del Zanna, L. Fernndez-Menchero, A. S. Giunta, G. Y. Liang, H. E. Mason, and P. J. Storey, J. Phys. B {\bf 49}, 
094001 (2016).

\bibitem{CC}
I. Bray, D. V. Fursa, A. S. Kheifets and A. T. Stelbovics, J. Phys. B \textbf{35}, 15 (2002).

\bibitem{Burke}
P. G. Burke, {\it R-Matrix Theory of Atomic Collisions}, Springer-Verlag Publication, Berlin (2013).

\bibitem{grasp2k}
P. J\"onsson, X. He, C. F. Fischer, and I. P. Grant, Comput. Phys. Commun. {\bf 177}, 597 (2007).

\bibitem{szabo}
A. Szabo and N. Ostuland, {\it Modern Quantum Chemistry}, Dover Publications, Inc., Mineola, New York, First edition(revised), 1996.

\bibitem{bartlett}
I. Shavitt and R. J. Bartlett, {\it Many-body methods in Chemistry and Physics}, Cambidge University Press, Cambridge, UK (2009).

\bibitem{plasma-rev}
R. K. Janev, S. Zhang, and J. Wang,
Matter Radiat. Extremes \textbf{1}, 237 (2016).

\bibitem{salzman}
D. Salzmann, {\it Atomic Physics in Hot Plasmas}, Oxford University Press, Oxford (1998).

\bibitem{Weisheit}
J. C. Weisheit,  Adv. At. Mol. Phys. \textbf{25}, 101 (1989).

\bibitem{Murillo}
M. S. Murillo, and J. C. Weisheit, Phys. Rep. \textbf{302} 1 (1998).

\bibitem{Saha}
B. Saha and S. Fritzsche, Phys. Rev. A, {\bf 73}, 036405 (2006).

\bibitem{Sil}
 A. N. Sil, J. Anton, S. Fritzsche, P. K. Mukherjee, and B. Fricke, Eur. Phys. J. D {\bf 55}, 645 (2009).

\bibitem{Thomas-Fermi}
Y. Jianmin, Phys. Rev. E \textbf{66}, 047401 (2002).

\bibitem{Ion-sphere-model}
S. Ichimaru, Rev. Mod. Phys. \textbf{54}, 1017 (1982).

\bibitem{sil2}
A. N. Sil, S. Canuto and P. K. Mukherjee, {\it Spectroscopy of Confined Atomic Systems: Effect of Plasma}, Advances in Quantum 
Chemistry, Chapter 4, pg. 115, Vol. 58 (2009). 

\bibitem{Zammit-1}
M. C. Zammit, D. V. Fursa, and I. Bray, Chem. Phys. \textbf{398}, 214 (2012).

\bibitem{Zammit-2}
M. C. Zammit, D. V. Fursa, and I. Bray, Chem. Phys. \textbf{82}, 052705 (2010).

\bibitem{Zhang-1}
S. B. Zhang, J. G. Wang and R. K. Janev, Phys. Rev. Lett. \textbf{104}, 023203 (2010).

\bibitem{Zhang-2}
S. B. Zhang,J. G. Wang, R. K. Janev  and X. J. Chen, Phys. Rev. A \textbf{83}, 032724 (2011).

\bibitem{Qi-1}
Y. Y. Qi, Y. Wu, J. G. Wang and Y. Z. Qu, Phys. Plasmas \textbf{16}, 023502 (2009).

\bibitem{Qi-2}
Y. Y. Qi, J. G. Wang and R. K. Janev, Phys. Rev. A \textbf{80}, 063404 (2009).

\bibitem{Ghoshal}
A. Ghoshal and Y. K. Ho, J. Phys. B \textbf{43}, 045203 (2010).

\bibitem{Chen}
Z. B. Chen, C. Z. Dong, J. Jiang  and L. Y. Xie, J. Phys. B \textbf{48}  144030 (2015).

\bibitem{bks1}
B. K. Sahoo, J. Phys. B {\bf 43}, 231001 (FTC) (2010).

\bibitem{bks2}
D. K. Nandy, S. Singh and B. K. Sahoo, MNRAS {\bf 452}, 2546 (2015).

\bibitem{bks3}
B. K. Sahoo and B. P. Das, Phys. Rev. A {\bf 92}, 052511 (2015).

\bibitem{bks4}
B. K. Sahoo, Phys. Rev. A {\bf 93}, 022503 (2016).

\bibitem{Mg+-importance-1}
E. Charro and I. Mart\'in,  Astrophys. J. \textbf{585}, 1191 (2003).

\bibitem{Mg+-importance-2}
 A. G. Jensen and T. P. Snow,  Astrophys. J. \textbf{669}, 401 (2007).

 \bibitem{Mg+-importance-3}
G. \c{C}elik, D. Do\u{g}an, \c{S}. Ate\c{s}, and M. Ta\c{s}er, J. Quant. Spectrosc. Radiat. Transfer \textbf{113}, 1601 (2012).

\bibitem{Mg-abundance-1}
N. F. Allard, G. Guillon, V. A. Alekseev and J. F. Kielkopf, A \& A \textbf{593}, A13 (2016).

\bibitem{Mg-abundance-2}
 M. Guitou, A. K. Belyaev, P. S. Barklem, A. Spielfiedel and N. Feautrier, J. Phys. B \textbf{44}, 035202 (2011).

\bibitem{LS-Mg+}
L. Sharma, A. Surzhykov, R. Srivastava, S. Fritzsche, Phys. Rev. A \textbf{83}, 062701 (2011).

\bibitem{Smith}
S. J. Smith, A. Chutjian, J. Mitroy, S. S. Tayal, R. J. W. Henry, K-F. Man, R. J. Mawhorter, and I. D. Williams,
Phys. Rev. A \textbf{48}, 292 (1993).

\bibitem{William-1}
I.~D. WIlliam, A.~Chutjian, and R.~J. Mawhorter, J. Phys. B \textbf{19}, 2189 (1986).

\bibitem{William-5CC}
I. D. Williams, A. Chutjian, A. Z. Msezane, and R. J. W. Henry, Astrophys. J. \textbf{299}, 1063 (1985).

\bibitem{YKKim}
Y.~K. Kim, Phys. Rev. A \textbf{65}, 022705 (2002).

\bibitem{Pangantiwar}
A.~W. Pangantiwar and R.~Srivastava, J. Phys. B \textbf{21}, L219 (1988).

\bibitem{BKS-1}
M. Das, B. K. Sahoo, and S. Pal,  Phys. Rev. A {\bf 93}, 052513 (2016).

\bibitem{BKS-2}
B. K. Sahoo and M. Das, Eur. Phys. J. D {\bf 70}, 270 (2016).

\bibitem{NIST}
[http://www.nist.gov/pml/data/asd.cfm].

\bibitem{grasp92}
F. A. Parpia, C. F. Fischer, and I. P. Grant, Comput. Phys. Commun. \textbf{94}, 249 (1996).

\bibitem{LosAlamos}
[http://aphysics2.lanl.gov/cgi-bin/ION/runlanl08d.pl].

\end{thebibliography}
\end{document}